\documentclass[pdflatex,sn-mathphys-num]{sn-jnl}

\usepackage{graphicx}%
\usepackage{multirow}%
\usepackage{amsmath,amssymb,amsfonts}%
\usepackage{amsthm}%
\usepackage{mathrsfs}%
\usepackage[title]{appendix}%
\usepackage{xcolor}%
\usepackage{textcomp}%
\usepackage{manyfoot}%
\usepackage{booktabs}%
\usepackage{algorithm}%
\usepackage{algorithmicx}%
\usepackage{algpseudocode}%
\usepackage{listings}%
\usepackage{ulem}
\usepackage{bm}
\usepackage{url}
\usepackage{subcaption}
\usepackage{multicol}

\newcommand{\D}[1]{} 
\begin{document}

\title[Article Title]{Dynamical evolution of social network polarization and its impact on the propagation of a virus}


\author*[1]{\fnm{Ixandra} \sur{Achitouv}}\email{ixandra.achitouv@cnrs.fr}

\author[1,2]{\fnm{David} \sur{Chavalarias}}

\affil*[1]{\orgdiv{Institut des Systèmes Complexes ISC-PIF }, \orgname{CNRS}, \orgaddress{\street{113 rue Nationale}, \city{Paris}, \postcode{75013}, \country{France}}}

\affil[2]{\orgdiv{Centre d'Analyse et de Mathématique sociales}, \orgname{CNRS}, \orgaddress{\street{54 boulevard Raspail}, \city{Paris}, \postcode{75016}, \country{France}}}


\abstract{The COVID-19 pandemic that emerged in 2020 has highlighted the complex interplay between vaccine hesitancy and societal polarization. In this study, we analyse the dynamical polarization within a social network as well as the network properties before and after a vaccine was made available. Our results show that as the network evolves from a less structured state to one with more clustered communities. Then using an agent-based modeling approach, we simulate the propagation of a virus in a polarized society by assigning vaccines to pro-vaccine individuals and none to the anti-vaccine individuals. We compare this propagation to the case where the same number of vaccines is distributed homogeneously across the population. In polarized networks, we observe a significantly more widespread diffusion of the virus, highlighting the importance of considering polarization for epidemic forecasting. }


\maketitle

\section{Introduction}

The COVID-19 pandemic that emerged in 2020 presented an unprecedented global health crisis. As governments and public health authorities raced to develop and distribute effective vaccines, a concerning trend emerged - the polarization of attitudes towards vaccination. 
This polarization was not merely a matter of individual choices, but rather reflected deep-seated divisions within societies. Numerous studies have documented the link between political ideology, media consumption, and COVID-19 vaccine hesitancy \cite{grabisch2021survey, rathore2020information, gisondi2022deadly}. In the USA, Republican sympathiser, for example, were found to be significantly more hesitant about getting vaccinated compared to Democrats, a divide that widened over the course of the pandemic \cite{rathore2020information}. This partisan gap in vaccine attitudes was exacerbated by the spread of misinformation, often amplified through ideologically-aligned media sources \cite{gisondi2022deadly, cascini2021attitudes}. Beyond political affiliation, vaccine hesitancy during COVID-19 also exhibited racial and socioeconomic disparities \cite{cascini2021attitudes, delgado2021covid}. Marginalized communities, who were disproportionately impacted by the pandemic, often expressed greater skepticism towards vaccines. This entanglement of vaccine hesitancy with societal polarization poses significant challenges for public health efforts. As the pandemic continues to evolve or in the event of a new global pandemic, understanding these complex dynamics is crucial for forecasting accurately the propagation of a virus and reducing its impact on society. 

In what follows, we use Twitter data related to COVID-19 vaccination and vaccine hesitancy over the years 2020 and 2022 as a proxy to study the dynamical polarization of society on this topic and identify online communities of people sharing similar opinions about vaccines. These online social networks are then used to simulate the propagation of a virus within networks with real-world properties \cite{Watts, leskovec2008statistical, mislove2007measurement} and to quantify the impact of polarization on epidemiological variables such as the cumulative attack rate. The key idea is that social structures matter, have a specific topology and properties and to consider that pro-vaccine communities will tend to have a good vaccination rate while anti-vaccines communities will have no vaccine coverage at all. We compare on networks with real-world properties this approach with the most common approach to epidemic modeling used for epidemiological forecast, where the vaccines are distributed homogeneously among the overall population (absence of structure on anti-vaccine individuals interactions). 

By using real social networks and an agent based model to simulate the transmission of a virus, we estimate how vaccine hesitancy, and the associated social polarization, influence the diffusion of a virus, which we find to be significantly different compared to the diffusion where we have an homogeneous vaccine coverage across the population. The collected annotated real-world networks are made available for the social science community as well as the software package developed for this work, see sec. \ref{secappend}.

This article is organized as follows: In sec.~\ref{secnetwork} we describe the datasets and the properties of the network. In sec.~\ref{ABM} we introduce the agent based model used to simulate the diffusion of a virus and in sec.~\ref{conclu} we discuss the main conclusion of this analysis.

\section{The network dynamics of pro-vaccines vs anti-vaccines}\label{secnetwork}
\subsection{Data collection and analysis of the communities}
From the start of the pandemic, we have monitored the Twitter messages that mentioned pandemic issues in relation to vaccination. 

Twitter data have been collected by means of the Twitter \textit{track API} between 2020 and 2022, which allows to capture all tweets mentioning a given expression. Over this period, we have  identified keywords and hashtags related to vaccination and filtered our original dataset to get a sub-dataset of 76,6M tweets containing at least one  keyword related to vaccination issues (see  sect.~\ref{dataCollection} for details). This vaccination related Twitter dataset, that spans from 2020-01-01 to 2022-12-31, contains among the 76,6M tweets, retweets with 12,3M users producing original tweets and 14,3M users having retweeted someone else at least once.

In \cite{gaumont_reconstruction_2018}, it is demonstrated that ideologically aligned groups could be retrieved as dense clusters of the retweet network, \textit{i.e.} groups of accounts that relay each other’s messages on a regular basis without modification. Filtering the retweet interactions (67,2M tweets), we build three retweets graphs over the years 2020, 2021 and 2022. 

To get as close as possible to the structure of real contact networks, we further filtered the retweet graph by selecting only two-way interactions, the assumption being that people retweeting each other are more likely to get to know each other offline and meet up physically. We thus got three bidirectional retweet graphs (cf. Table~\ref{tab:tab1}). Following  \cite{gaumont_reconstruction_2018},  we have identified the epistemic communities of those three networks using the Louvain clustering algorithm \cite{blondel_fast_2008}. We then manually assessed the attitude towards vaccination of all communities larger than 1\% by reviewing the content produced by their main influencers on their Twitter account and removed the communities that were not clearly positioned in favor or against vaccination. Influencers have been identified both as accounts with the highest pagerank and accounts having produced the highest combined number of tweets and retweets. As a result, we obtained three real-world networks made up of users with two-way links and a clear position in favor or against vaccination.

\begin{figure}
    \centering
	\begin{subfigure}{.45\textwidth}
		\includegraphics[width=\textwidth]{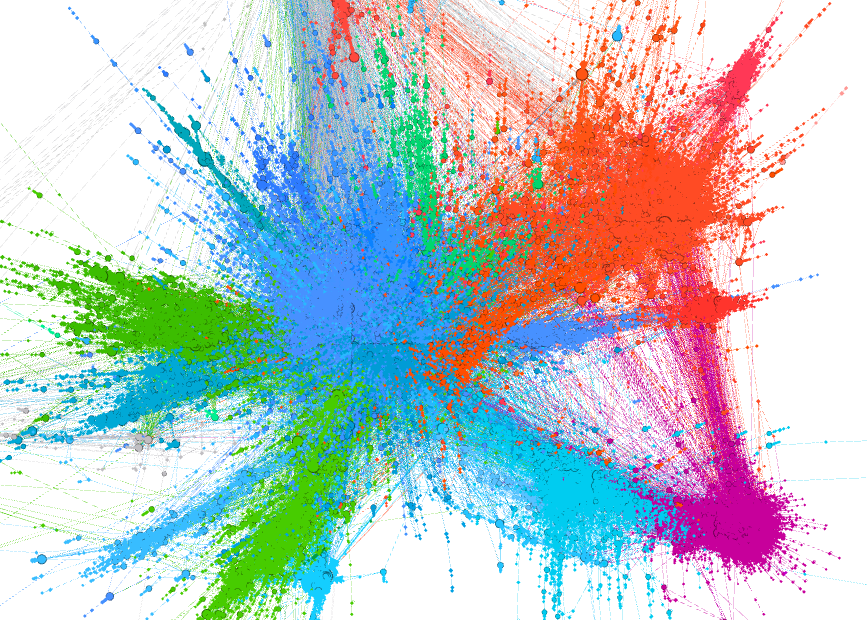}
		\caption{2020}
	\end{subfigure}
	\begin{subfigure}{.45\textwidth}
		\includegraphics[width=\textwidth]{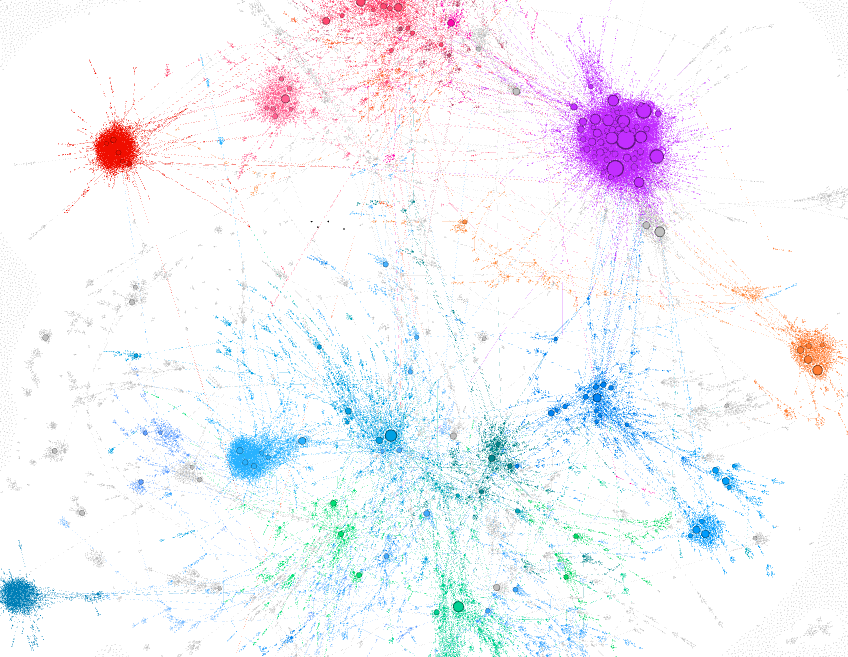}
		\caption{2021}
	\end{subfigure}
	\begin{subfigure}{.45\textwidth}
		\includegraphics[width=\textwidth]{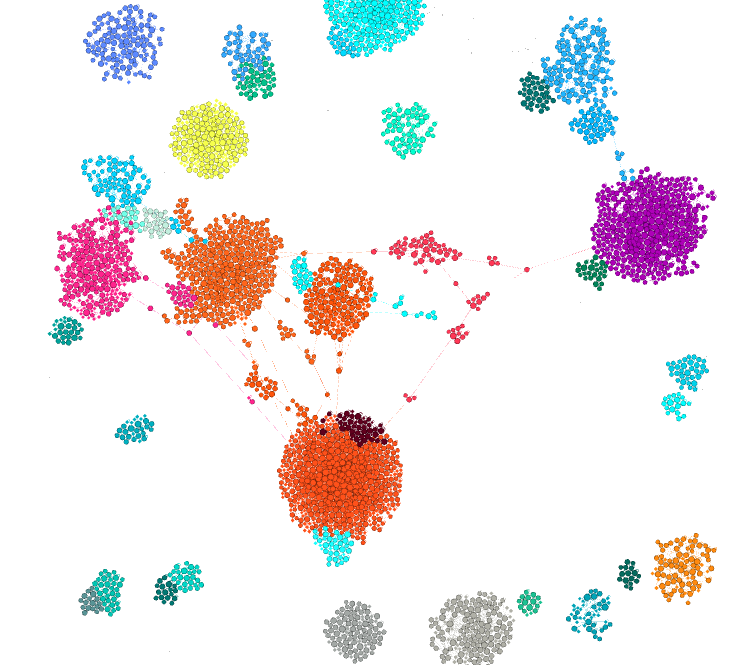}
		\caption{2022}
	\end{subfigure}
	\caption{Visualization of social network on bidirectional retweets among users sharing COVID related tweets. Cold colors correspond to pro-vaccines users and warm colors to anti-vaccines users. Visualization have been generated with the Gephi software \cite{gephi}.}

    \label{fig:network}
\end{figure}

\subsection{Network properties analysis}

\begin{table}
    \centering
    \begin{tabular}{|c|c|c|c|c|c|}
    \hline
       year  & users  & edges  & $\%$ anti-vaccines \\
       \hline
        2020 & 113038 & 223099 & 29 \\
        2021 & 18826 & 75750 & 47 \\
        2022 & 3617 & 14604 & 69 \\
    \hline
    \end{tabular}
    \caption{network properties evolution of COVID related topics for bidirectional tweets.}
    \label{tab:tab1}
\end{table}

In order to quantify the dynamical evolution of our networks, we analyzed their global structural similarities and differences by measuring their density, their clustering coefficient (CC) and the degree Distribution. We also compare the difference of these measure within the pro and anti-vaccines communities. 

\begin{figure}
    \centering
    \includegraphics[width=0.7\linewidth]{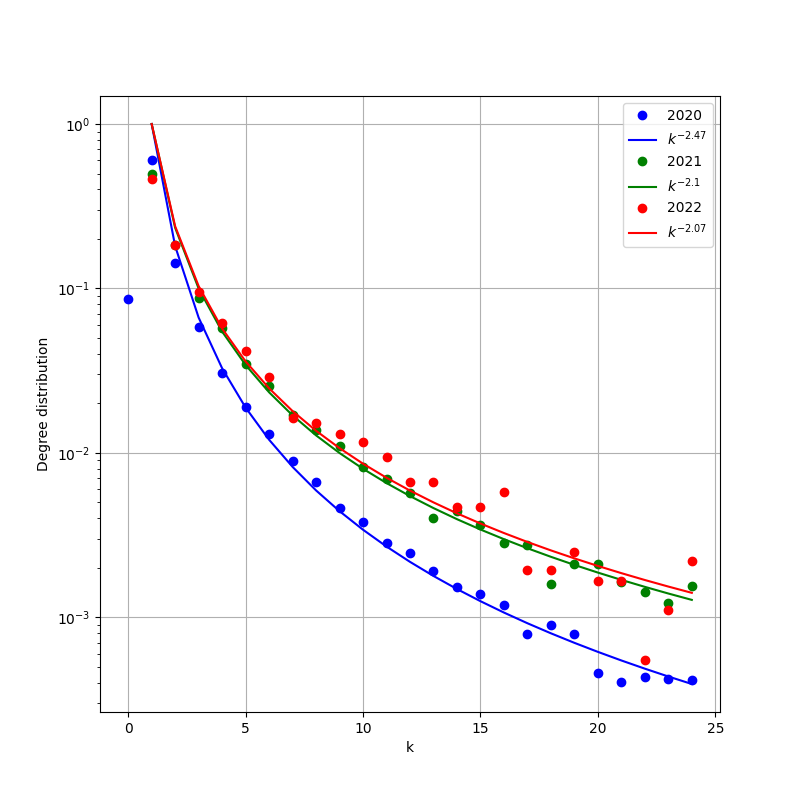}
    \caption{Networks degree distribution and the associated fit in $P(k)\sim k^{-\gamma}$  }
    \label{fig:deg-PL}
\end{figure}

The summary is given in Tab.\ref{tab:tab2} and the degree distribution is displayed in Fig.\ref{fig:deg-PL}. In Tab.\ref{tab:tab2}
we also report the mean number of daily contacts which increases in time here mainly because in 2020 there were many users without strong views discussing the pandemic while in 2022 the number of users have reduced, leaving users with strong opinions. 

\begin{table}
    \centering
    \begin{tabular}{|c|c|c|c|c|c|c|}
    \hline
       year  & Density  & CC & A & $\bar{I}$\\
       \hline
        2020 & 0.00003  & 0.05 & 0.92 & 1.95\\
        2021 & 0.00043  & 0.17 & 0.99 & 2.\\
        2022 & 0.00223  & 0.18 & 0.99 & 3.7\\
    \hline
    \end{tabular}
    \caption{Network global properties evolution of covid related topics for bidirectional tweets, where A is the assortativity and $\bar{I}$ is the mean number of daily interactions.}
    \label{tab:tab2}
\end{table}

In what follows, a graph (or network) $G=(V,E)$ formally consists of a set of vertices (or nodes) $V$ and a set of edges $E$ between them. An edge $e_{ij}$ connects vertex $v_{i}$ with vertex $v_{j}$. 

\subsubsection{Density}

Network density measures how many edges are present in a network compared to the maximum possible number of edges and is defined as: 
\begin{equation}
    D=\frac{E}{E_{max}}
\end{equation}
where $E$ is the number of edges, $E_{max}$ is the maximum number of edges in our network and is given by $n(n-1)/2$ with $n$ the total number of nodes. It is a measure of how interconnected a network is. In the context of a social network like Twitter, density quantifies the proportion of actual connections out of all possible connections between users. The 2022 network has the highest density, indicating a more interconnected structure where information can spread more easily through the network, while 2020 network is the sparsest with isolated clusters and less efficient information flow. Interestingly one may compute the density within pro-vaccine and anti-vaccine communities, see Tab.\ref{tab:tab3}.  

 \begin{table}
    \centering
    \begin{tabular}{|c|c|c|}
    \hline
       year  & Density pro-vaccines & Density anti-vaccines\\
       \hline
        2020 & 0.00004  & 0.00016 \\
        2021 & 0.00067  & 0.00109 \\
        2022 & 0.00529  & 0.00360\\
    \hline
    \end{tabular}
    \caption{Networks density evolution among the pro-vaccines vs anti-vaccines communities.}
    \label{tab:tab3}
\end{table}

In 2020 and 2021 we observe that the anti-vaccine community has a density which is an order of magnitude larger than within the pro-vaccines community, pointing out that the anti-vaccines community is already a well-cohesive community with many exchanges compared to the pro-vaccine individuals.

\subsubsection{Degree Distribution}

The degree of a node $i$ in a network is the number of connections or edges the node has to other nodes $j$ with $i\neq j$. The degree distribution P(k) of a network is then defined to be the fraction of nodes in the network with degree k. Hence for a network with $n$ nodes where $n_k$ of them have degree $k$, we have:
\begin{equation}
     P(k)=\frac {n_{k}}{n}
\end{equation}
   
The degree distribution is a property that characterizes the connectivity patterns in a network. For social networks like Twitter, it shows the probability distribution of how many connections (here bidirectional retweets) users tend to have.

The degree distribution of social networks often follows a power-law shape, meaning the probability of a node having k connections decreases as a power of k: $P(k)\sim k^{-\gamma}$ where $\gamma$ is a positive parameter. This indicates a scale-free topology with no characteristic degree scale where few nodes are highly connected (decreasing in power of $\gamma$). This is fundamentally different than in the case of the Erdos-Renyi random network \cite{Erdos} where the degree distribution follows a Poisson distribution. Note that  some social networks are small world but not always scale free (e.g. \cite{zachary_karate_1977}, \cite{guimera_self-similar_2003}). In Fig.\ref{fig:deg-PL} on can see the degree distribution evolution as a function of  year (colored dots) as well as a power law fit (solid colored curves) where $\gamma\sim 2-2.5$ similarly to previous studies e.g. \cite{e17085848}. 
Among the pro and anti-vaccines communities, the degree distribution also follows a power law where the exponent is reported in Tab.\ref{tab:tab4bis}. 

\begin{table}
    \centering
    \begin{tabular}{|c|c|c|c|}
    \hline
       year  & $\gamma$ global & $\gamma$ pro-vaccines & $\gamma$ anti-vaccines\\
       \hline
        2020 & 2.47 &  2.64  & 2.22 \\
        2021 & 2.1 & 2.14  & 2.06 \\
        2022 & 2.07 & 2.18  & 1.97\\
    \hline
    \end{tabular}
    \caption{Networks degree distribution fit among the pro and anti-vaccine communities.}
    \label{tab:tab4bis}
\end{table}
These fitting parameters indicate that within the anti-vaccine community the connections of users are higher than in the pro-vaccines communities with a larger number of highly connected users.

\subsubsection{Clustering Coefficient (CC)}

The clustering coefficient is a measure of the degree to which nodes in a graph tend to cluster together. In social networks, nodes tend to cluster \cite{Watts}. The local clustering coefficient of a node $v_i$ in an undirected graph is 
\begin{equation}
   C_{i}=2\frac {|\{e_{jk}:v_{j},v_{k}\in N_{i},e_{jk}\in E\}|}{k_{i}(k_{i}-1)}, 
\end{equation}
which is the ratio between the cardinal of the node $v_i$ neighbourhood (immediate connected neighbours), while $k_i(k_i-1)/2$ is the number of links that could possibly exist between node $v_i$ and its neighbours $k_i$. Thus the average clustering coefficient over all nodes $n$ is given by:
\begin{equation}
   \bar{CC}={\frac {1}{n}}\sum _{i=1}^{n}C_{i}.
\end{equation}

As reported in Tab.\ref{tab:tab2} the $\bar{CC}$ increases as time evolves. This can be interpreted as an increase of the polarization: user opinion becoming more clustered within their community. Here again we can compute the clustering coefficient among the pro and anti-vaccine communities displayed in Tab.\ref{tab:tab4}. 
 \begin{table}
    \centering
    \begin{tabular}{|c|c|c|}
    \hline
       year  & CC pro-vaccines & CC anti-vaccines\\
       \hline
        2020 & 0.04  & 0.07 \\
        2021 & 0.18  & 0.16 \\
        2022 & 0.19  & 0.18\\
    \hline
    \end{tabular}
    \caption{Networks clustering coefficient among the pro-vaccines vs anti-vaccines communities.}
    \label{tab:tab4}
\end{table}

Moreover, there is a significant difference in the clustering coefficient of the pro and anti-vaccine communities in 2020, anti-vaccines being almost twice as clustered as pro-vaccines which is in line with their highest density and underline the strong community structure of these groups.

\subsubsection{Assortative mixing in networks}\label{Assor}
In order to characterize the degree of clustering among pro and anti-vaccine individuals, we also consider the associativity measure given by:
\begin{equation}
r=\frac{Tr \; \textbf{e}-\lVert \textbf{e}^2 \rVert}{1-\lVert \textbf{e}^2 \rVert}
\end{equation}

where $\textbf{e}$ is a 2x2 matrix whose elements are $<e_{i,j}>$ (the average is over all individuals) and the index $i,j$ can have the values ${0,1}$ for unvaccinated and vaccinated individuals, respectively. Thus $e_{1,1}$ corresponds to the fraction of connections a vaccinated agent has with other vaccinated agents, $e_{1,0}$ the fraction of connections a vaccinated agent has with unvaccinated agents and $e_{0,0}$ the fraction an unvaccinated agent has with other unvaccinated agents.  For an undirected network, $\textbf{e}$ is symmetric.

With this definition, $-1\leq r \leq 1$ where the limits $r=\left\lbrace 1,0,1\right\rbrace $ correspond to completely anti-correlated, random and completely correlated groups \cite{Newman}. In Table~\ref{tab:tab1} we report that the polarization of our network is high and increasing from 2020 to 2021 and 2022.

Interestingly one can also quantify the evolution of exchanges between the pro and anti-vaccine communities by considering the off-diagonal terms $e_{1,0}$. However due to the difference in the ratio of pro and anti vaccines users we consider instead a measure that quantifies the cross connections between our two communities: $2 e_{1,0}/(e_{1,1}+e_{0,0})=0.0111, 0.0009, 0.0006$ for 2020, 2021 and 2022 respectively. This means that at the start of the epidemic, the two communities were communicating $\sim 20$ times more than in 2022. 

\subsection{Sociological interpretation}
Analyzing the network properties on a time scale that spans the pre-covid and the post-covid vaccines shows an increasing evolution of the polarization between users (assortativity in tab.\ref{tab:tab1}) as well as intrinsic evolution of the network properties. The evolution of the  clustering coefficient demonstrate that the probability that a link exists between two arbitrary neighbors of an arbitrary node increases as well as the efficacy to diffuse information (density), but reduce when two nodes are in the opposite community (cross-connections decrease). The network evolves from a less structured network to one with more connections among more clustered communities. Interestingly the networks do follow a power law as expected from small world network.

The network topology plays a major role in information propagation \cite{nekovee2007theory, wang2003epidemic, Perra2015, goncalves2012social}. Hence one might wonder how the diffusion/transmission of ideas or a virus, is affected by polarization. In what follows, using our networks as toy models to capture polarization in society, we apply a probability of transmission of a virus to study its propagation across users. We consider the covid transmission probability as an example but our key results focus on how the transmission is affected by polarization in the case of a major event such as the novelty of a vaccine. While there have been many studies on the impact of the network in the transmission of a disease e.g. \cite{sahasranaman2020spread,Salathe, Fisman, Nielsen} or in societal segregation \cite{Grossmann} based on vaccination status \cite{Hickey2022.08.21.22279035, Achitouv2022}, this work compares directly the differences of the propagation in a segregation based on vaccination status (polarization) through time using the interactions of a real social network which differ from most networks used in such modeling: Erdos–Rényi, Watts-Strogatz and Barabási–Albert models.

In the Erdos–Rényi random network \cite{Erdos} there is no community structure nor power law in the degree of distribution. In the Watts-Strogatz model \cite{Watts}, the generated network produces small word properties including short average path lengths and high clustering. The probability of connecting two nodes is not random and can be tuned. However this type of network does not generate a degree distribution that follows a power law observed in many real world networks \cite{Newman}. It is interesting to point out that in \cite{Hinch, Achitouv2022} the social network of work interactions generated for propagating the virus is in fact a Watts-Strogatz model. In the Barabási–Albert model \cite{Albert_2002}, the generated network produces the power law distribution of the degree distribution as well as short average path lengths but fails to produce the high levels of clustering seen in real networks. Therefore, these two last models are particularly useful to generate networks but cannot be considered as fully realistic.

In complex sciences, dynamics of an opinion and its diffusion in a social group have been largely investigated \cite{castellano2009statistical, friedkin1986formal, degroot1974reaching, proskurnikov2018tutorial, grabisch2021survey, Albert_2002}. Thus, it is particularly relevant to use real-world networks for investigating how the diffusion of a virus (or opinions) is affected by polarization compared to the diffusion in an unpolarized network. In what follows, we describe the model we build and how we apply it to the social networks we collected in sec~\ref{secnetwork}. 

\section{Impact of polarization in the diffusion of a virus}\label{ABM}

\subsection{The agent based model and transmission probability}
Simulations serve a crucial function in exploring social dynamics. They mirror the conventional methods employed in theoretical physics, where the objective is to characterize a system through a set of equations. These equations are subsequently solved either numerically, through computational techniques, or analytically, whenever feasible, to gain insights into the system's behavior \cite{Castellano_2009}.

One of the most successful methodologies employed in the study of social dynamics is agent-based modeling \cite{schweitzer2003brownian}. These include pedestrian traffic modeling, simulations of urban aggregation processes, and the study of opinion formation dynamics. The core idea behind this approach is to construct computational entities, known as agents, imbued with specific properties, and then simulate their parallel interactions to model real-world phenomena. In these models, individual agents (which may exhibit heterogeneity) interact within a given environment, adhering to procedural rules governed by characteristic parameters. Agent-based simulations have emerged as a powerful tool across various scientific disciplines, particularly in the exploration of social systems see the seminal work by \cite{epstein1996growing}.
 
 In this work, we develop an agent based model with a structural network taken from on the online social networks we described previously and with infection characteristics of COVID-19 as a study case. Every day we loop over all contacts of the infected individuals and we compute the probability of infecting non-infected individuals from their contact lists. We consider that the probability of infection (see~\cite{Hinch} and reference within) is given by:

\begin{equation}
P(t,s,a,n)=1-\exp[-\lambda(t,s_i,a_s,n)], 
\end{equation}
with $\lambda$ is the rate at which the virus is transmitted in a single interaction:
\begin{equation}
\lambda(t,s_i,a_s,n)=\frac{R S_{a_s} A_{s_i} B_n }{\bar{I}} \int_{t-1}^{t} f_{\Gamma}(u;\mu_i;\sigma^{2}_{i}) du ,\\
\end{equation}

where $t$ is the time since infection, $a_s$ indicates the infector's symptom status (asymtomatic, mild, moderate/severe), $s_i$ is the age of the susceptible and $n$ is the type of network where the interaction occurred. We only consider one network hence $n=1,B_n=1$ and we consider a case where the infectious rate $R=4$. 
The denominator $\bar{I}$ is the mean number of daily interactions, in our case it is computed from the distribution of bidirectional retweets, see Table\ref{tab:tab1}. In the integral, $f_{\Gamma}(u;\mu;\sigma^{2})$ is the probability density function of a gamma distribution with $\mu=5.5$(days) and $\sigma=2.14$ the mean and width of the infectiousness curve see \cite{Ferretti2020},\cite{He2020}. $R$ scales the overall infection rate, $S_{a_s}$ is the scale-factor for the age of the susceptible and $A_{s_i}$ is the scale-factor for the infector being asymptomatic. In our study we assume no knowledge on an individual's age and symptom status. Thus we average over the values given in \cite{Hinch} leading to $A_{s_i}=0.88$, $S_{a_s}=1.14$.

\subsection{The vaccine scheme}
In this analysis we consider that all pro-vaccine individuals received a complete vaccine scheme and none of the anti-vaccine individuals received vaccines.

When an individual is infected and vaccinated, we loop over his contact list only if $u>VET$ where u is a uniform random number between $\left[ 0,1\right[ $ and VET is the vaccine effectiveness against transmission. We consider $VET=0.9$ which corresponds to the alpha SARS-COV-2 estimate from \cite{PMID:37085456}. When looping over the contact list of the infected individual, the infection on a vaccinated individual is reduced and is computed only if $v>VEI$ where $v$ is a uniform random number between $\left[ 0,1\right[$ and $VEI=0.6$ is the vaccine effectiveness based on the transmission probability for infectiousness \cite{Madewell2022}.

\subsection{Results}
\begin{figure}
    \centering
	\begin{subfigure}{.99\textwidth}
		\includegraphics[width=\textwidth]{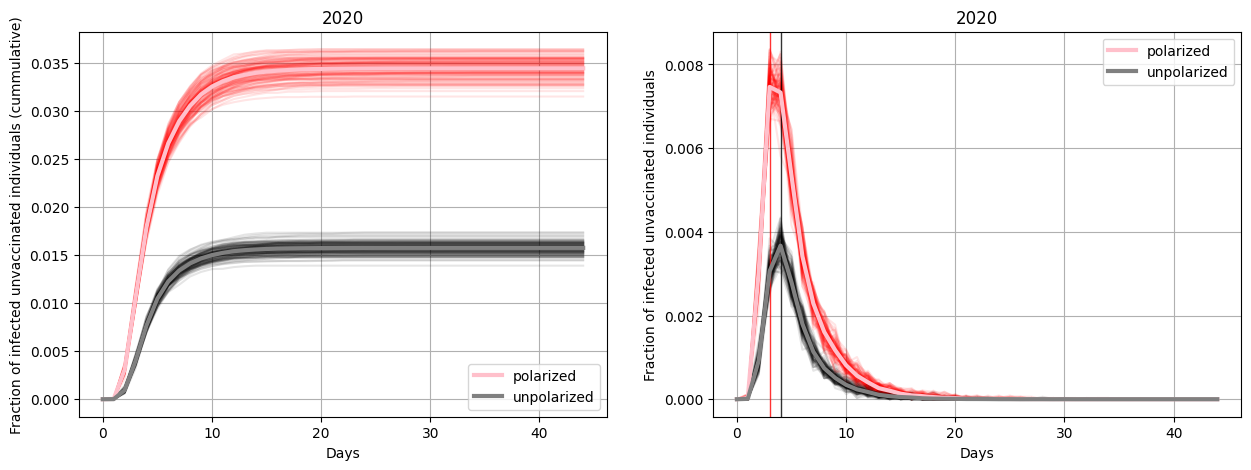}
	\end{subfigure}\\
	\begin{subfigure}{.99\textwidth}
		\includegraphics[width=\textwidth]{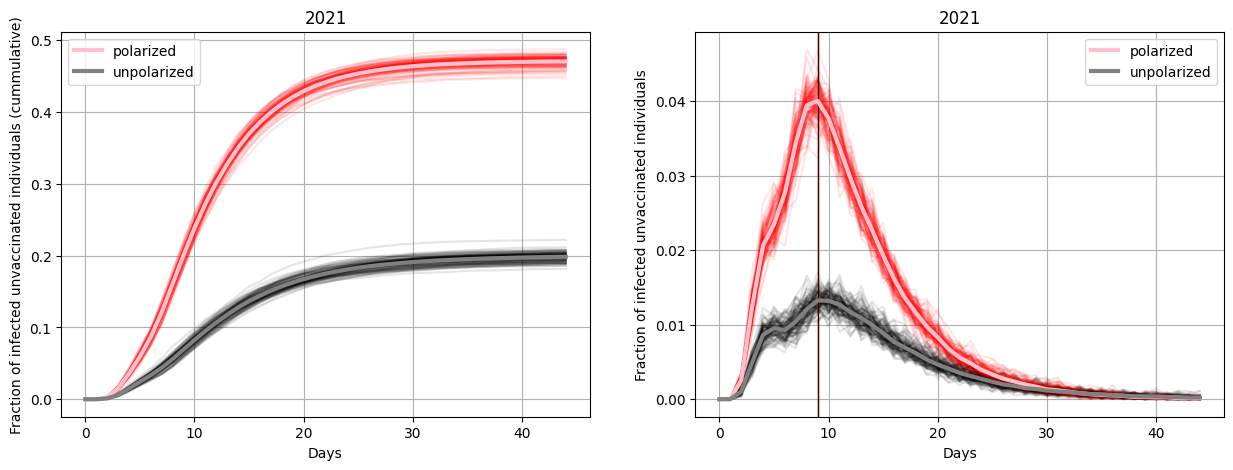}
	\end{subfigure}\\
	\begin{subfigure}{.99\textwidth}
		\includegraphics[width=\textwidth]{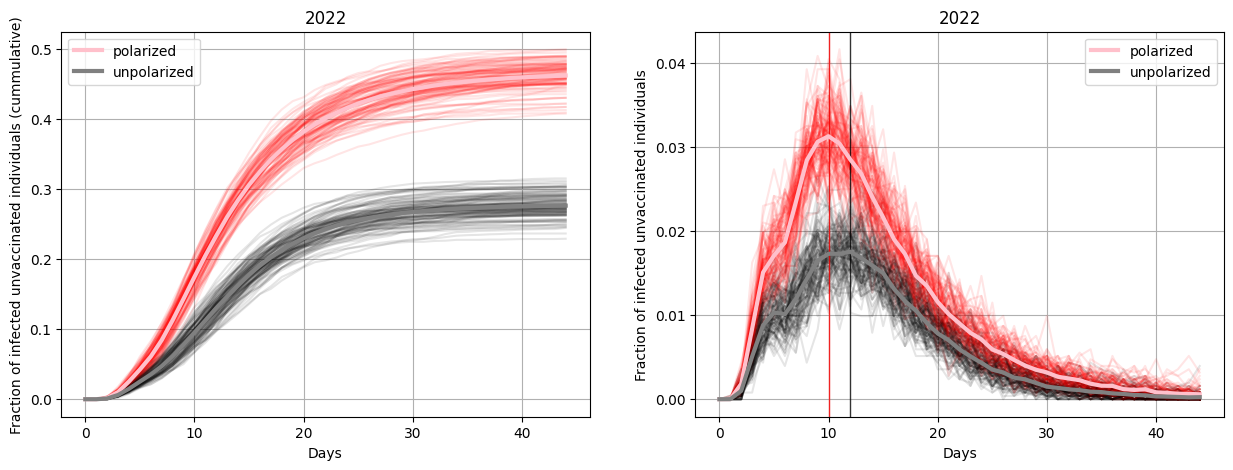}
	\end{subfigure}
	\caption{Polarization impact on epidemic dynamics: fraction of daily number of infections (right panels) and the cumulative fraction of infection (left panels)}
    \label{fig:Ninf}
\end{figure}

For each network of Tab. \ref{tab:tab1} we run 100 simulations. To quantify the impact of polarization over a random attribution of the vaccines across all individuals, we run in each network another 100 simulations by taking the same number of vaccines but allocating them uniformly across all individuals such that the assortativity is null. The results are displayed in Fig.\ref{fig:Ninf} where the red curves correspond to the polarized attribution of the vaccines and the grey curves correspond to a random attribution. 

These curves show the infection curves within the unvaccinated individuals normalized by the number of unvaccinated individuals which is the population at risk. In appendix \ref{Appen_inf} one could see the outcomes across all the population and among the vaccinated population. 

\begin{table}
    \centering
    \begin{tabular}{|c|c|c|c|c|}
    \hline
       year  & AR polarized  & AR unpolarized  & $T_{peak}$ polarized  & $T_{peak}$ unpolarized \\
       \hline
        2020 & 3$\%$ & 1$\%$ & 3 & 4 \\
        2021 & 47$\%$ & 20$\%$ & 9 & 9 \\
        2022 & 46$\%$ & 27$\%$ & 10 & 11 \\
    \hline
    \end{tabular}
    \caption{summary of the epidemic properties for the polarized vs unpolarized network among the unvaccinated individuals}
    \label{tab:tab3inf}
\end{table}

The key results are summarized in Tab.\ref{tab:tab3inf}. The attack rate (AR) is defined as the cumulative fraction of individuals who were contaminated at the end of the simulation and $T_{peak}$ is the time (in days) at which the number of daily contaminations is maximum. Unsurprisingly, the diffusion of the virus within a polarized network is faster and impacts more individuals than in an unpolarized society. The overall increase ratios are $2.4,2.4,1.7$ for the years 2020, 2021 and 2022 respectively. This can be interpreted as a screening mechanism, when vaccinated individuals are mixed with the unvaccinated ones they block the diffusion of the virus while in an unpolarized society where the vaccines are distributed homogeneously, there is no such screening among the unvaccinated individuals. 

Hence, our results demonstrate that the diffusion of the virus within these polarized network structures is faster and more widespread compared to unpolarized social landscapes. The tight-knit communities acted as efficient conduits for disease transmission, leading to approximately twice the expected number of contaminated individuals.

\section{Discussion}\label{conclu}

The polarization of society is an increasing area of research within developed economies ~\cite{3wiki}. For instance, it has been well studied in wealth inequalities for more than a few decades, but the polarization of opinion is a phenomenon that can only be observed recently at large scales using social media \cite{bail2018exposure, barbera2015birds, hemphill2016polarscores, garimella2018quantifying, calais2013measure}. The COVID-19 pandemic has laid bare the profound impact that the evolution of social networks and the increasing polarization of societies can have on public health outcomes. This study demonstrates that these intertwined dynamics can exacerbate the spread of infectious diseases.

Firstly, we observed a concerning shift in the structure of social networks over the course of the pandemic. The social fabric was torn apart as the epidemic unfolded (see sec.~\ref{Assor}). The initially less organized network gradually transformed into one characterized by more interconnected, tightly-clustered communities. This transition likely reflects the human tendency to seek out ideologically-aligned social circles, especially during times of crisis and uncertainty. Compounding this structural change, we also found that societal polarization has intensified, with a growing percentage of individuals expressing anti-vaccine sentiments (Tab.\ref{tab:tab1}), a decrease in the exchanges between pro and anti-vaccine users (sec.~\ref{Assor}) and denser connections between like-minded individuals (Tab.~\ref{tab:tab2} \& \ref{tab:tab3}). We also report that the communities between pro and anti-user accounts are structured differently (see Tab.~\ref{tab:tab3}, Tab.~\ref{tab:tab4} and \ref{tab:tab4bis}) where the anti-vaccines seems to be more cohesive (denser and higher clustering coefficient). One could draw a parallel from an epidemiology point of view: not applying preventive measures and social distancing among anti-vaccine individuals naturally leads in higher interactions within this community.

Secondly, the results of this research have significant implications in policy-making and public health as we find that for a polarized society structured in clusters, the spread of the virus, for a given vaccine coverage of the population, is significantly different to the case where vaccines are distributed homogeneously. Thereby it has a negative impact on the pandemic control by (i) accelerating disease spread in the community --which can quickly overwhelm hospital capacity-- and (ii) increasing the final cumulative attack rate, approximately twice as much as in the homogeneous case. In particular, it goes against one of anti-vaccine campaigners' favorite arguments that there was no point in vaccinating, since the vaccine neither prevented the virus from being caught nor transmitted, and therefore had no impact on the management of the epidemic. Not only does vaccine hesitancy accelerate the spread of the virus, saturating hospitals, but the polarizing attitude of anti-vaccinators amplifies this phenomenon. Therefore, understanding and mitigating the polarization within societies need to be addressed in order to control the course of a global epidemic. 

Thirdly, we provide in this work a new dataset from a real social network annotated by pro or anti-vaccine opinion as well as the agent based simulation used to simulate the diffusion of a COVID-19 like virus. This material will be useful for future research to investigate different local rules for the evolution of agent states. For instance, one could test the evolution of opinion of users who are connected to both communities. One could also analyze situations where there are less available vaccines than people willing to get them, with a polarized society on top of this. This would make it possible to better control the experiment independently of the structure of clusters and might be closer to reality at the start of  pandemics.

To conclude, our study highlights the importance of considering opinion dynamics alongside disease dynamics, providing insights into the interplay between propensity to follow heath recommendations, opinions dynamics, and disease outcomes. The associated networks and model contribute to understand the impact of societal polarization in the propagation of a virus.

\backmatter


\bmhead{Acknowledgements}
We warmly thank the ISC-PIF multivac platform and Maziyar Panahi for the data collection of the tweets and Jeremy Ward for his help in the identification of vaccine related keywords. 

\bmhead{Funding}
This work was supported by the Complex Systems Institute of Paris Île-de-France (ISC-PIF), the Région Île-de-France SESAME 2021 and the EU Nodes project (Grant Agreement LC-01967516).

\bmhead{Author contribution}
IA and DC designed the research study and wrote the
paper. IA programmed the original ABM simulation and performed the quantitative analysis of the networks. DC conducted the empirical analysis on the networks annotating the opinion of the communities. 

\bmhead{Data availability}
The Twitter dataset is not available for public release because it violates the Terms of Service and contains identifying information. An anonymous dataset is available at \url{https://github.com/IxandraAchitouv/ABM-social-network-.git}

\bmhead{Code availability}
Code availability The code to reproduce the analysis in this paper is available at \url{https://github.com/IxandraAchitouv/ABM-social-network-.git}

\section*{Declarations}
\bmhead{Conflict of interest}
On behalf of all authors, the corresponding author states that there is no conflict of interest

\bmhead{Ethics approval}
`Not applicable'


\begin{appendices}

\section{Appendices}\label{secappend}
\subsection{Software package and anonymous networks}

The agent based simulation for this work as well as the social networks annotated (pro or anti-vaccines) are being made available at \url{https://github.com/IxandraAchitouv/ABM-social-network-.git}.
The Id of users are labelled randomly at a given year. Hence id $i$ in 2020 is not the same user as $i$ in 2021. 
If one is interested in following the opinion of a user please contact us. 

\subsection{Data collection \label{dataCollection}}
Twitter data have been collected by means of the Twitter \textit{track API} between 2020 and 2022, which allows to capture all tweets mentioning a given expression. 

We have collected all the tweets associated to the following keywords : 
\begin{multicols}{3}\footnotesize 
\begin{itemize}
\item coronavirus
\item covid-19
\item covid\_19
\item covid19
\item covid2019
\item confinement
\item restezchezvous
\item chloroquine
\item restezàlamaison
\item covid2019france
\item jerestechezmoi
\item PJLcoronavirus
\item masques
\item covidfrance
\item covidfr
\item covid19fr
\item coronavirusfrance
\item coronavirusfr
\item déconfinement
\item Coronavirus
\item Koronavirus
\item Corona
\item CDC
\item Wuhancoronavirus
\item Wuhanlockdown
\item Ncov
\item Wuhan
\item N95
\item Kungflu
\item Epidemic
\item outbreak
\item Sinophobia
\item China
\item covid-19
\item corona virus
\item covid
\item covid19
\item sars-cov-2
\item COVID-19
\item COVD
\item pandemic
\item coronapocalypse
\item canceleverything
\item Coronials
\item SocialDistancingNow
\item Social Distancing
\item SocialDistancing
\item panicbuy
\item panic buy
\item panicbuying
\item panic buying
\item 14DayQuarantine
\item DuringMy14DayQuarantine
\item panic shop
\item panic shopping
\item panicshop
\item InMyQuarantineSurvivalKit
\item panic-buy
\item panic-shop
\item coronakindness
\item quarantinelife
\item chinese virus
\item chinesevirus
\item stayhomechallenge
\item stay home challenge
\item sflockdown
\item DontBeASpreader
\item lockdown
\item lock down
\item shelteringinplace
\item sheltering in place
\item staysafestayhome
\item stay safe stay home
\item trumppandemic
\item trump pandemic
\item flattenthecurve
\item flatten the curve
\item china virus
\item chinavirus
\item quarentinelife
\item PPEshortage
\item saferathome
\item stayathome
\item stay at home
\item stay home
\item stayhome
\item GetMePPE
\item covidiot
\item epitwitter
\item pandemie
\end{itemize}
\end{multicols}

We have analyzed most keywords and hashtags related to vaccination and created a sub-dataset of 76.6M tweets containing at least one  of these keywords: 
\begin{multicols}{3}\footnotesize 
\begin{itemize}
\item vaccin
\item vaccins
\item arnm
\item arn messager
\item hpv
\item gardasil
\item zostavax
\item dtvax
\item agrippal
\item avaxim
\item vaqta
\item havrix
\item bexsero
\item boostrixtetra
\item cervarix
\item dukoral
\item encepur
\item engerix
\item fluarix
\item fluarixtetra
\item fluenz
\item genhevac b pasteur
\item hbvaxpro
\item hexyon
\item immugrip
\item imovax polio
\item infanrixquinta
\item infanrix
\item infanrixtetra
\item dtpca
\item influvac
\item ixiaro
\item rvaxpro
\item meningitec
\item menjugate
\item menjugatekit
\item menveo
\item mosquirix
\item menbvac
\item neisvac
\item nimenrix
\item pentavac
\item pneumo
\item pneumovax
\item prevenar
\item priorix
\item rabipur
\item repevax
\item revaxis
\item rotarix
\item rouvax
\item rotateq
\item spirolept
\item stamaril
\item tetravac
\item ticovac
\item twinrix
\item tyavax
\item typherix
\item typhim vi
\item meningococcique
\item varilrix
\item varivax
\item vaxigrip
\item zostavax
\item pentavalent
\item hexavalent
\item heptavalent
\item monovalent
\item trivalent
\item squalène
\item thiomersal
\item thimerosal
\item focetria
\item pandemrix
\item panenza
\item celvapan
\item humenza
\item vaccination
\item vaccinations
\item adjuvant
\item adjuvante
\item adjuvantes
\item antivaccin
\item antivaccins
\item anti-vaccin
\item antivaccinaliste
\item antivaccinalistes
\item antivaccinationiste
\item antivaccinationniste
\item antivaccinationistes
\item antivaccinationnistes
\item vaxxer
\item vaxxers
\item antivax
\item antivaxer
\item antivaxers
\item antivaxxer
\item antivaxxers
\item vaccinal
\item vaccinaux
\item vaccinale
\item vaccinales
\item vaccin
\item vaccins
\item pfizer-biontech
\item moderna
\item astrazeneca
\end{itemize}
\end{multicols}

This dataset was used to produced online Twitter interaction network related to vaccination.

\subsection{Additional infection curves}\label{Appen_inf}

\subsubsection{Infection curves among vaccinated individuals}

\begin{figure}[h]
    \centering
	\begin{subfigure}{.99\textwidth}
		\includegraphics[width=\textwidth]{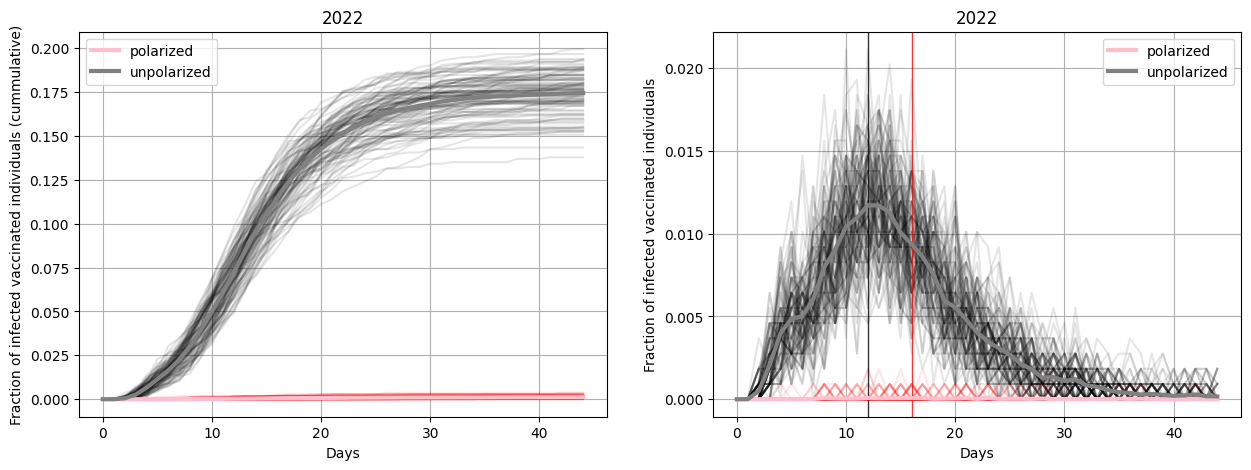}
	\end{subfigure}\\
	\begin{subfigure}{.99\textwidth}
		\includegraphics[width=\textwidth]{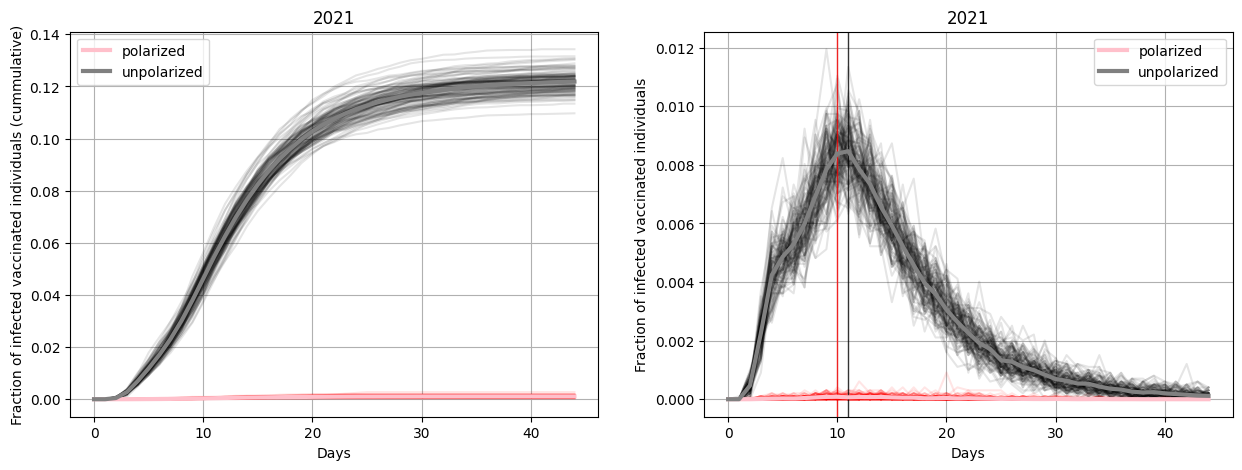}
	\end{subfigure}\\
	\begin{subfigure}{.99\textwidth}
		\includegraphics[width=\textwidth]{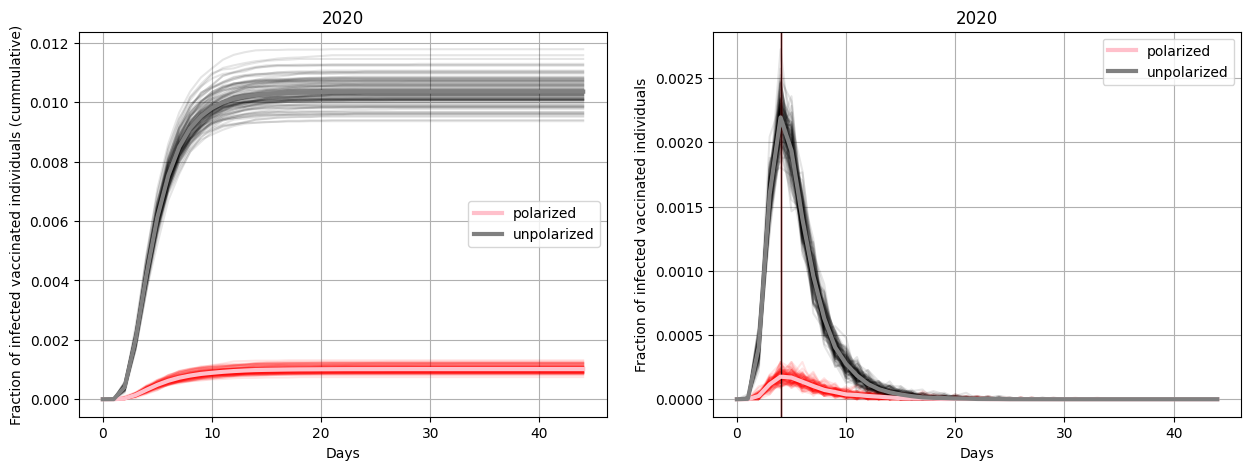}
		
	\end{subfigure}
	\caption{Polarization impact on epidemic dynamics: fraction of daily number of infections (right panels) and the cumulative fraction of infection (left panels) among vaccinated individuals}

    \label{fig:Ninfvacc}
\end{figure}
Within the vaccinated population, the polarization acts as a barrier toward contamination. It is the opposite behaviour among the unvaccinated individuals. The probability of transmitting the virus is reduced when vaccinated hence the polarization decreases the propagation of the epidemic among cluster of vaccinated individuals which is not the case in an unpolarized society see Fig.\ref{fig:Ninfvacc}.

\subsubsection{Infection curves among all individuals}
\begin{figure}[h]
    \centering
	\begin{subfigure}{.99\textwidth}
		\includegraphics[width=\textwidth]{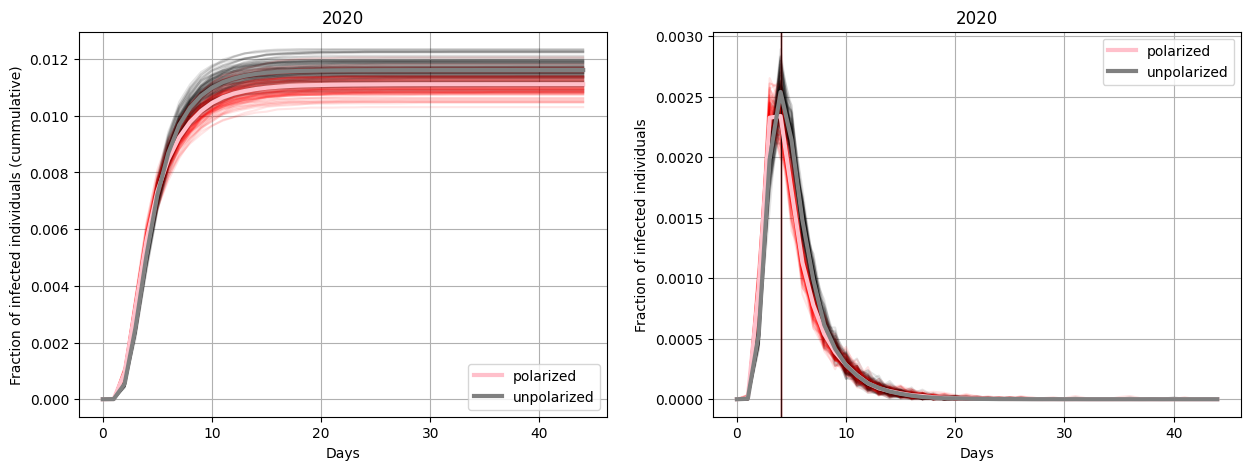}
	\end{subfigure}\\
	\begin{subfigure}{.99\textwidth}
		\includegraphics[width=\textwidth]{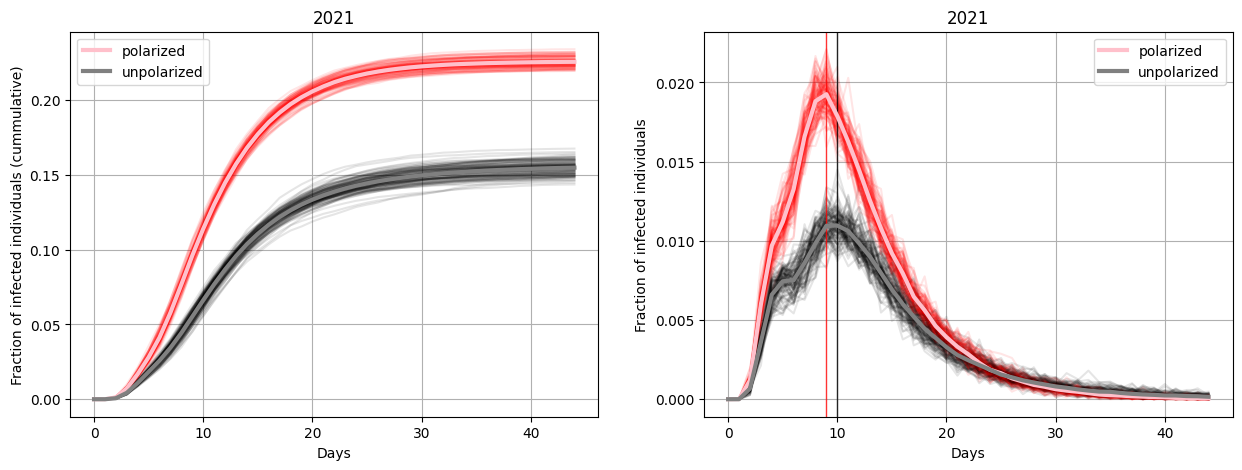}
	\end{subfigure}\\
	\begin{subfigure}{.99\textwidth}
		\includegraphics[width=\textwidth]{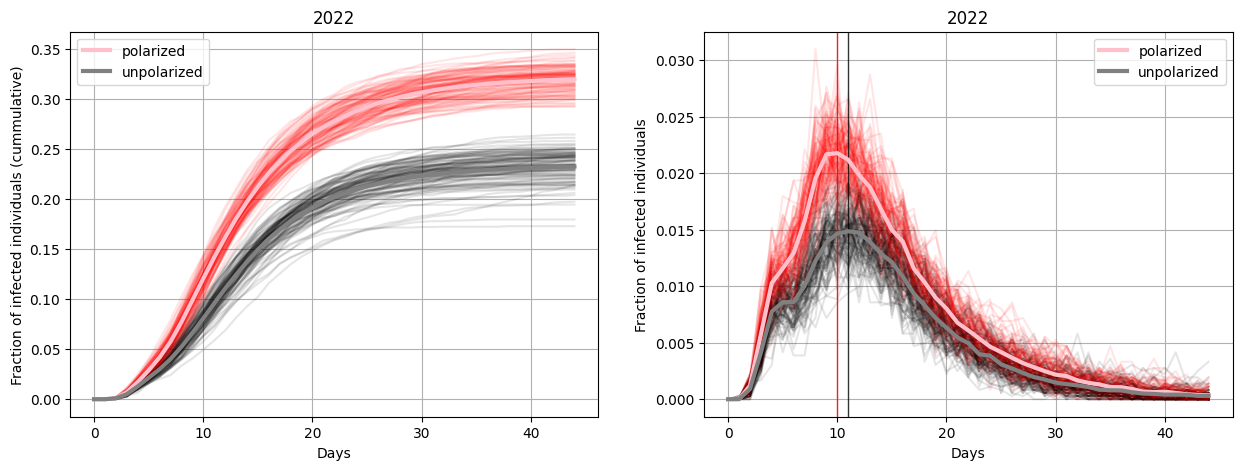}
	\end{subfigure}
	\caption{Polarization impact on epidemic dynamics: fraction of daily number of infections (right panels) and the cumulative fraction of infection (left panels) among all individuals}

    \label{fig:Ninfall}
\end{figure}

When considering all individuals (vaccinated and unvaccinated), the effect of polarization is reduced because of the two opposites effects: the propagation is reduced among vaccinated individuals and it is enhanced among the unvaccinated individuals see Fig.\ref{fig:Ninfall}.

\end{appendices}


\bibliography{sn-bibliography}

\end{document}